\documentclass[twoside, openright]{nictatechreport}

\usepackage{latexsym,amsmath,amssymb,amsfonts}
\usepackage{url}
\usepackage{multirow}
\usepackage[english]{babel} 

\usepackage{graphicx}
\graphicspath{{./images/}}

\usepackage{enumerate}

\newenvironment{compact_itemize}
  {\begin{itemize}\addtolength{\itemsep}{-.5\baselineskip}}
  {\end{itemize}}

\usepackage[usenames]{color}
\definecolor{MyDarkBlue}{rgb}{0,0.08,0.7}

\title{A Case for a Global Information Network}
\author{Maximilian Ott$^1$, Yan Shvartzshnaider$^2$}

\otheraffiliation{$^1$Corresponding author: max.ott@nicta.com.au\\ $^2$Also with University of Sydney}

\submitdate{NetDB 2011, April 1, 2011}
\reportnumber{TR-4783}
\pubhistory{Submitted to NetDB 2011}
\headertitle{Ott, Shvartzshnaider, A Case for a Global Information Network}

\begin{document}

\frontmatter
\pagestyle{empty}
\begin{abstract}


This paper argues for the adoption of a information centric system model instead of the current service-oriented one. We present an architecture for a global information storage and dissemination network which provides for efficient interaction and coordination among autonomous actors through a shared information space. We believe that the resulting, loosely coupled systems, while probabilistic in nature, will lead to robust outcomes at large scales.



%

\end{abstract}





\mainbody
\pagestyle{fancy}

\newpage
\newlength{\picturewidth}
\setlength{\picturewidth}{0.75\columnwidth}

\section{Motivation}\label{motivation}

There is an ever increasing gap between how the ``common man'' would describe the Internet and how a network researcher would. This is not surprising, as the ``packet with a label identifying the receiver'' model seems to be an ill fitting abstraction for  popular services provided by companies such as Google, Facebook, YouTube, or Twitter. These services are primarily about information dissemination and a suitable network service abstraction should really do more than deliver individual packets with minimal latency. But how would such a service look like? 

The web taught us that much information is in the links between ``things``. Services, such as email, chat, and voice mail, clearly demonstrate that most of our interactions spread across time as well. The database community has clearly embraced the challenge of ``linked data at unprecedented scale'' as reflected by new research on graph or no-schema databases, as well as different distributed models. On the other hand, the networking community has not much progressed beyond a slightly more generalised notion of packet and time in the form of content delivery networks. 

\begin{figure}[b]
\centering
	\includegraphics[width=\picturewidth]{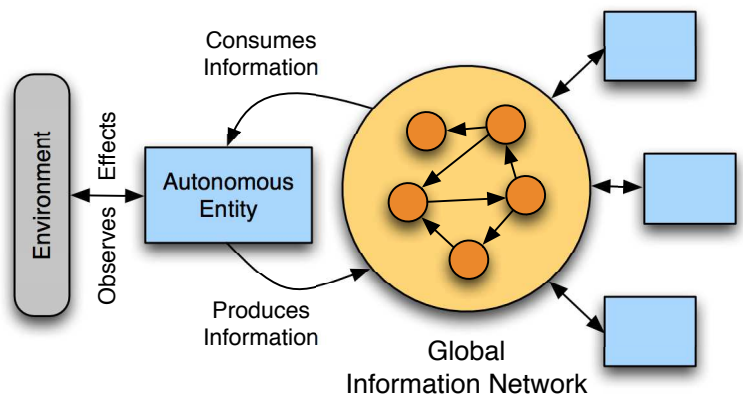}
\caption{Information Centric System Model.}
\label{model}
\end{figure}

So far, this has been primarily about supporting the delivery of specific, well defined services to a potentially large number of users. If a specific service, such as email, is not provided by a single company, we have defined standards, such as SMTP, to govern information exchange within that domain. But otherwise there was little incentive or willingness to expose the information contained within a service. But this is starting to change, an increasing number of popular services, such as Facebook and Twitter, are providing programatic interfaces to a user's information, leading to an array of interesting ``mesh-up`` services.

In the reminder of this paper we will argue that we should adopt an information centric system model instead of the current service oriented one. Instead of dealing with information exchange between specific services on a case by case bases, we should instead consider a global information storage and dissemination network onto individual applications and services are built. Information exchange is done implicitly through sharing of it as depicted in Figure~\ref{model}.

So what are the high-level design goals for such a new network service?
\begin{compact_itemize}
\item Support for linked data
\item Scalability
\item Usability
\item Trustworthyness
\end{compact_itemize}

As mentioned above information is fundamentally about capturing the properties of entities and the relations between them. The lack of a uniform way of supporting this simple abstraction across our applications and services lead to unnecessary complications and a plethora of special purpose exchange points and formats. Just think about all the complications to keep messages, contacts, and calendar consistent if one doesn't stay within the confine of a single provider. The Semantic Web community has assembled and defined many of the necessary components, but it is lacking a broader service model.

Adding scalability to the design goals of any network service is almost redundant. However, a global information network is going to add a few additional magnitudes to the challenge as any device can produce many nodes and especially links to other nodes which can lead to substantial changes to the topology of the information graph.

For any new service abstraction to have a chance of being adopted it not only needs to provide sufficiently new useful functionality but it's service model needs to fit the model of the service or application built on top. Many commonly used design methodologies, such as object-oriented have an inherent graph structure and the many existing middleware services, mapping from and to object graphs, further validate the usability of a graph abstraction.

Finally, any Internet-scale service abstraction needs to contain some fundamental building blocks from which to bootstrap trustworthy services. We chose trust over security as a design goal as we do not believe that the inherently black \& white world of security is realistic in the context of large systems. However, it is unclear if there is a widely accepted definition of what makes a system trustworthy without needing to consider external factors. We therefore are focusing on securing the provenance of the information added to the graph, which we believe is sufficient to bootstrap other mechanisms that may establish the trustworthiness of some information within a specific context.

In the reminder of this paper we will first briefly discuss related work, then describe the proposed architecture in more depth, followed by an overview of our related prototyping activities and close with a conclusion.

\label{motivation}
\section{Related Work}

The proposed information network is a cumulation of many ideas from the fields of networking, database and the semantic web. This section is highlighting some of them.

\textbf{Network.} Finding alternatives to packet switched networks which better fit the emerging use cases and design methodologies has been an active field of research for many years. We are especially interested in systems which only loosely couple the sender and receiver. 

In publish-subscribe (P/S) systems the subscriber defines what published ``events'' to receive. Widely deployed P/S systems, which are now often referred to as Enterprise Bus, provide ``channels'' as rendezvous abstraction, or subscribe with a set of constraints on events structured as lists of key/value pairs. Eugster, et al, provide a comprehensive overview in\cite{Eugster:2003:MFP:857076.857078}. However, these systems are usually limited to a specific organisation or service and are realised through a (federated) broker architecture. 

There have been various attempts to scale P/S systems, with most of them based on a ``rendezvous'' model. For instance, PSIRP \cite{fotiou_illustratingpublish-subscribe_2009} only uses the rendezvous to facilitate the maintenance of more efficient delivery paths.

In contrast Jacobson, et al, treat content as a primitive in their Content-Centric Networking  (CCN) system\cite{jacobson2009networking}. Hierarchical names which identify the content source are used to router {\em interest} packets toward that source. Intermediate node who may have cached that content can consume the {\em interest} and send the content data block towards the requester. Furthermore,  forwarding nodes on that path may decide to cache the content as well. An embedded security model as well as a pull operation address many of the threat scenarios of the current Internet.




\textbf{Database.} The explosion of data behind many popular online services has forced the database community to adapt and reinvent itself to provide a better, more flexible service for the application and network layers.  Furthermore, new business models behind today's Internet systems have led to the development of large-scale storage systems such as Dynamo\cite{decandia_dynamo:_2007}, BigTable\cite{chang2008bigtable} and Cassandra\cite{website:cassandra}, with associated processing capabilities, such as Hadoop \cite{website:hadoop}. These systems, often referred to as NoSQL systems,  relax the traditional ACID properties, to provide a fully distributed, highly available, partition tolerant storage.   

Another side of this evolution is the emergence of "graph information representation" that captures more naturally the diverse and heterogeneous nature of data in today's applications\cite{eifrem2009neo4j}.  Graph databases such as Neo4J\cite{eifrem2009neo4j} and others offer fully transactional database system - storing information as a graph rather than tables. 

The scalability properties behind NoSQL systems and the effective manner in which data can be captured and mined using a graph structure has led  to increased efforts to combine the two approaches to facilitate fully distributed, and large-scale graph data processing. Cloudera uses map/reduce techniques to efficiently analyse large-scale graphs\cite{website:cloudera}. 
Recently, Malewicz, et al introduced 
Pregel\cite{malewicz2010pregel},  a vertex-centric framework for efficient 
processing of large-scale graphs. A user provided program is periodically executed in the context of each individual vertex. Messages sent to a downstream vertex enter the context of that vertex in the next, globally synchronised, iteration cycle, called ``superstep''. Overall processing terminates when all vertex contexts have terminated.


\textbf{Semantic Web.} The semantic web community has developed many mature standards and mechanisms, such as RDF\cite{w3c:rdf} to describe information, and OWL\cite{w3c:owl} to describe the ontology governing a specific information domain. With the W3C as the main standards body for many of these activities, the primary use case is the Web.

Fensel\cite{fensel_triple-space_2004} introduced the idea of a triple-space as a communication and coordination framework for the Semantic Web and Semantic Web services, where communication is done by writing to and reading from a shared, persistent triple space. Fensel's triple spaces are essentially a specialisation of tuple spaces used in parallel-processing models, such as Linda\cite{Gelernter:1992:CLS:129630.129635}.

\label{related_work}
\section{Architecture}

In this section we present our architecture in a top-down manner starting with enumerating the design goals.

\textbf{``Reflectable'' graph abstraction.} While we argued above that graphs are a natural way to describe the information inherent in a service, it is often necessary to capture ``information about information''. In the context of graphs with named edges, this is often referred to as ``4-hypergraph''. 

\textbf{Semantics as an end-to-end property.} We need to cleanly separate the fundamental abstraction of {\em named} vertices and edges from the specific semantic meaning of such a name.  In other words, we want to find the smallest set of definitions and operations provided by the network service which will still allow us to describe rich information spaces governed by evolving ontologies.
 
\textbf{Asynchronous communication.} For many use cases the time between information creation and consumption can be arbitrarily large and there seems to be little value in providing two distinct services, one for storage and one for (instantaneous) communication. 

\textbf{Time-stamped, insert only.} In a world where growth in storage capacity continues to substantially outperform Moore's law, it seems silly to ever throw anything away. While certain information may no longer be valid, there is still value in knowing that it was at some time, or at least someone believed it to be valid at a certain time. Whereas current legal requirements mandate the ability to remove information, it is practically impossible to do that for any document which was accessible on the web for some time. However, this does mean that time needs to become a fundamental attribute of any piece of information. 

\textbf{Best effort service.} One can argue that the ``best effort'' service model of the Internet is one of the key reasons it could grow to its current scale. End-to-end mechanisms, such as TCP together with proper traffic engineering lead to mostly acceptable service performance. And we see similar mechanism employed in very large storage systems through proper balancing of redundancy and storage medium reliability.

\textbf{Scoped trusted contexts.} How do we maintain privacy in a global information network? We can use the information graph itself to annotate parts of the graph, which may represent sensitive information, with information on who can access it and under which conditions. But who will be enforcing this? A pragmatic approach is to have multiple service providers, similar to ASes in the current Internet, which maintain the part of the global network created by their respective users. Global queries are being forwarded to each other and executed on the local graph in accordance to the associated privacy settings. This is clearly the weakest part of our case, but our architecture does not introduce fundamentally new security concerns, it simple makes them more apparent.


\textbf{Deployability.} One of the key challenges for any new networking service is to reach critical mass. Fortunately, the ubiquity of IP connectivity and abundance of cloud storage makes it relatively easy to start such a service as an overlay which can easily be joined by any Internet connected device. In fact, the real challenge is to be able to transparently deploy new resources and system upgrades as the service grows and matures.

\textbf{Scope.} The final key decision refers to the scope of such a service. Should it be able to do everything the current Internet does as well? While this is the starting assumption for many ``Future Internet'' initiatives, it seems prudent to take a more minimalistic approach and initially delegate the realisation of certain functionality to other existing systems. An example of that are nodes representing blobs, such as images or videos which are already well served by content-centric networks.


\subsection{Architecture Realization}

The system model depicted in Figure~\ref{model} is very similar to a publish-subscribe one but with a few interesting twists. Normally, a publication is a self contained event which is matched against all the active subscriptions. In our case, a publication persistently extends the global graph, while a subscription is essentially a standing query on the entire graph, or a specific mapping of the global graph to a local one. 


However, the two basic operations supported by our system, are {\em publish} and {\em subscribe}. But before we look at them in some detail, we first need to define the mechanism to describe the graph or parts of it. 

\subsubsection{Graph Representation}

Vertices in our directed graph either represent a specific instance of a ``thing'' or a ``value''. In fact, there is only exactly one vertex per instance in the graph. Values are an array of bits with a type specification whose interpretation is an end-to-end concern. However, a standard set of types is understood by the network to provide a broader query capability, as we will see later on.

\begin{figure}[!ht]
\centering
	\includegraphics[width=\picturewidth]{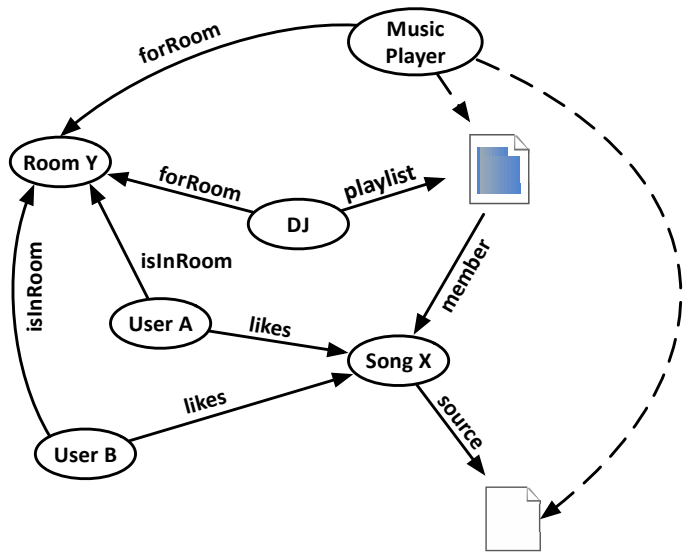}
\caption{Sample Information Network.}
\label{packet_structure}
\end{figure}

Edges on the other hand, represent a specific relationship between two nodes and are defined by those two nodes and a label identifying the type of relationship. Note, the semantic meaning of the label remains undefined here, however two edges with the same label are assumed to share the same semantic interpretation. In other words, while nodes are labeled uniquely, many edges may have the same label.

With these definitions in hand, we can now describe the entire graph or any portion of it as a list of tuples, each comprised of the label of the ``from'' vertex, the edge label, and either the label of the ``to'' vertex, or the value/type pair of a value node. To cover some of the above stated design goals we also add a ``context'' label and a time stamp, and optionally a ``signer'' label, and the hash of the entire tuple encrypted by the creators private key.

The context label provides a reflection capability as it allows us to represent a sub graph (tuples with the same context) as a vertex in the graph which can be linked to other vertices and values in the same way as any other vertex. While we have not settled on the trust \& security aspect of the architecture, we currently support basic provenance tracking by signing each tuple. Whereas there are various alternative options, we do want to point out that we can use the graph itself to maintain most of the necessary meta information associated with a PKI system. The signer is represented by a vertex in the graph with a relation to a value vertex containing its respective public key with that association signed by the next node in the trust chain. White and black listings can easily be captured through relations to the respective vertices. 

The labels themselves have very few requirements to fulfill; the need to be unique and it should be easy to establish if two labels are identical or not. Due to the nature and purpose of the graph there is no inherent value in encoding some form of hierarchical structure in the label and we therefore chose  UUIDs\cite{rfc4122} as our representation for labels. The fixed width of a label leads to a tuple serialisation as depicted in Figure~\ref{packet_structure}. This format is very similar to a conventional network packet, with the header containing all the fixed-width components and an optional payload containing the value of a value node. Please note, that there is no use case for an edge between two value nodes and value nodes are restricted to be on the receiving side of the directed edge from a ``labeled'' node.

\begin{figure}[!ht]
\centering
	\includegraphics[width=\picturewidth]{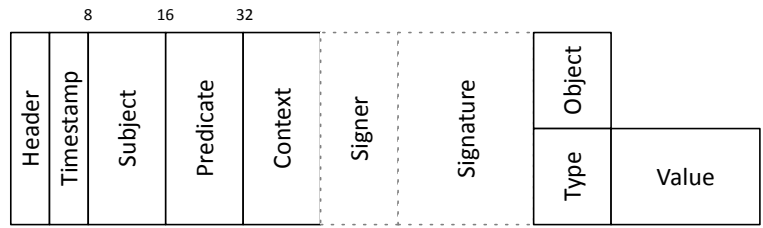}
\caption{Packet structure.}
\label{packet_structure}
\end{figure}

\subsubsection{Publish}

The {\em publish} API only needs to support the ``sending'' of a sequence of tuples similarly to a raw socket interface. Depending on the deployment of the overall service it will most likely be necessary to provide some additional context to be able to connect, but there is no fundamental need for any session or target context. 

\subsubsection{Subscribe}

The {\em subscribe} API is slightly more complex as every subscription creates a stateful context. A subscribe {\em session} is associated with an {\em interest} description and will result in an incoming stream of tuples. While there are many graph query languages available, for the moment, we settled at the simplest possible one onto which we can layer implementations of more powerful ones. Specifically, we define a subscription as a filter template  on all tuples currently available in the graph and subsequently added during the time the session is active. This solution requires all joins to be performed at the end devices. 

This simple filter mechanism clearly has performance implications concerning the number of tuples which will get dropped in subsequent joins as well as the lack of flow control on the receiving end. There are many solutions in the literature to address these problems but our initial design objective was to keep it as simple as possible.

\label{architecture}
\section{Prototype Implementation}

In order to better understand the implications of our design goals, we have implemented various parts of an overall system, including applications, higher-level subscription languages, and implementations of the network service as well. In the reminder of this section, we will briefly touch on each of these areas.

\subsection{Room DJ}

\begin{figure}[b]
\centering
  \includegraphics[width=\picturewidth]{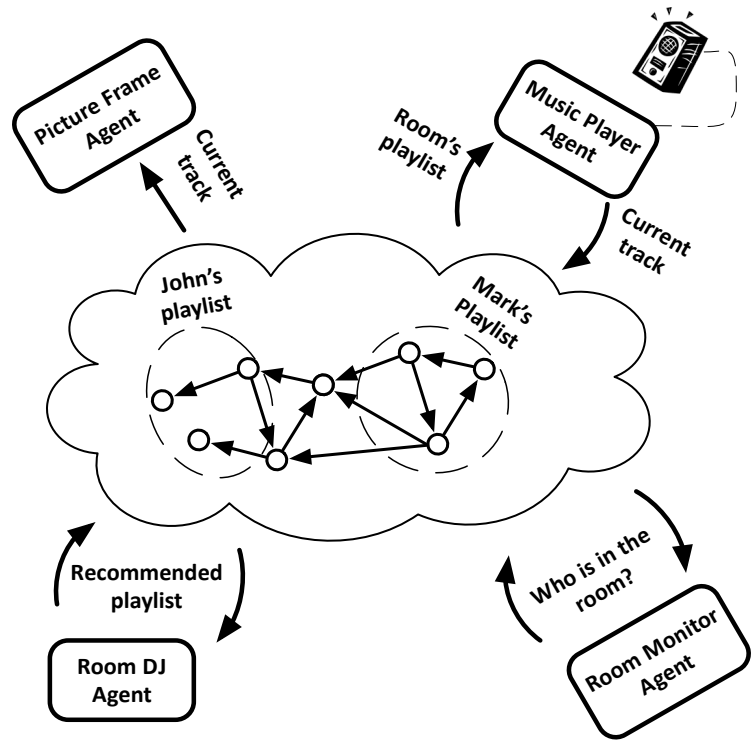}
\caption{Room DJ Setup.}
\label{sdj_model}
\end{figure}

To demonstrate the advantage of an information network centric model, we have implemented the following simple demo scenario. In order to simplify things, we assume that somehow a ``likes`` relationship between users and various songs are maintained in the system, as well as ``usesMobilePhone'' and the respective MAC address of that phone's WiFi radio. We also instrumented a room with a WiFi monitoring station, a network connected loudspeaker, as well as picture frame and built a few simple, independent agents. The ``Room Monitor`` agent is periodically pulling the list of observed MAC addresses from the monitoring station, queries for the associated phone and its user and finally publishes ``userX is in roomY'' tuples. The ``Room DJ'' agent subscribes to the ``all song liked by all users in roomY'' sub graph and periodically publishes a recommended playlist. The ``MusicPlayer'' agents subscribes to that playlist and the ``contentURL'' of the respective songs and instructs the network speakers what to play. It also publishes ``songZ is playing in roomY'' facts. Finally, the ``Picture Frame'' agent  is subscribing to ``the artwork of the song which is playing in roomY'' and displays that on its screen. 

The last agent demonstrates the motivation for the timestamp associated with each tuple. There will be potentially many tuples stating that different songs are playing in that room, the time stamp indicates when each statement was true (or believed to be true by someone). Although this slightly complicates the subscription query of the picture frame agent as it is only interested in what is happening right now, however it does provide other agents with  access to the entire history for whatever data mining purpose they sought. The room monitor agent also exposes an interesting aspect related to ``crossing the subscription boundaries''. Whereas it would be easy for the room monitor to maintain state and publish a ``user X is no longer in room Y'' there are many cases where it is much harder to capture a state change which is indirectly manifested by the publication of a new event which does not fall within a subscription. For instance, let us assume that each mobile phone in our scenario is periodically publishing its current geo-location and we know the geo-position of a room. The room monitor could now subscribe to ``any person whose location is within the room Y's boundary'' which would keep it informed whenever a new person enters the room, but not necessarily when that person leaves. One solution could be to temporarily subscribe to the location of any detected person until they report a location outside the room. However, we believe that a soft-state approach based on periodic publications of current state with corresponding time-outs on the consuming side most likely lead to more robuster solutions and are worthwhile the overhead.

\subsection{RETE}

Influenced by the Internet's hour glass design objective, we attempted to define the thinnest waist possible on the interface between the service provider and service consumer realm. In this section, we describe a more powerful subscribe language and it's implementation.

The above described subscribe operation only defines a mechanism to subscribe at the tuple level. Subscribing at the graph level requires performing joins across tuple streams. Figure~\ref{sample_subscription} depicts a subscription from the ``Room DJ'' agent implementation that we described above where the '?' identifies binding variables.

\begin{figure}[t]
\centering
\begin{minipage}{0.5\columnwidth}
\begin{verbatim}
SUBSCRIBE ?s
WHERE
  ?p isIn ?r.
  ?r name 'Lab A'.
  ?p likesSong ?s
\end{verbatim}
\end{minipage}
\caption{Sample Subscription}
\label{sample_subscription}
\end{figure}

To realise this capability in our context we have been adapting Rete\cite{doorenbos_production_1995}, an efficient pattern matching algorithm frequently used in production rule systems. Rete generates a dataflow network from a given set of rules to process a flow of tuples\footnote{As it is not relevant for our context, we ignore RETE's ability to also support the removal of tuples}. These dataflow networks consist of alpha memory (AM), beta memory (BM) and join vertices (JV). The basic topology pattern of a RETE network is a JV downstream from exactly one AM and BM feeding into another BM. Each AM is associated with a predicate and conceptually stores all received tuples which matched that predicate. Whenever an AM receives a new tuple, it sends a {\em right activation} event to its downstream JV. The JV in turn performs a join operation between that tuple and the set maintained by the upstream BM. Any newly discovered bindings are turned into tuples and pushed to the downstream BM. That BM in turn sends  {\em left activation} events for any newly received tuple to its downstream JV which reacts by performing a join operation with the current set maintained by its AM. Any newly discovered bindings are treated the same way as described above for the right activation event.

In our implementation, each AM is initially associated with a low-level subscription. For instance, the query in Figure~\ref{sample_subscription} translates to three AMs, with the following subscriptions: [$*$, isIn, $*$], [$*$, name, 'Lab A'], and [$*$, likesSong, $*$] with the labels ``isIn'', ``name'', and ``likesSong'', replaced by their respective UUIDs.

Obviously, basic query patterns with multiple wildcards can results in a large number of matches. However, our experience so far indicates that we can usually re-arrange the Rete network to team a ``large'' AM with a small BM. In this case, we can replace the single broad subscription with a number of more selective subscriptions. We believe that the vast literature on joins and query planning will provide us with many more ideas to optimise further.

\subsection{P2P-based Network Implementation}

The flat naming scheme and an initial overlay deployment strategy invite the consideration of many of the available P2P network solutions. Whereas P2P addressing is one-dimensional, various solutions have been reported to deal with the multi-dimensionality of tuples. RDFPeers\cite{rdfPeers} stores each 3-tuple three times, using each individual tuple element as a separate key. Besides the enforced redundancy, this scheme violates the uniform distribution assumption in the key space. However, it requires no modification to the routing protocols. 

Our prototype builds on the Kademlia protocol\cite{kademlia} which we extended in the following way. To simplify the explanation we will drop support for value vertices and only consider a network with labeled vertices and edges. Hence every tuple (ignoring the time stamp, signer and signature) translates into a 4 $*$ 128 bit long address. Since the 4 individual segments represent specific entities and relations, we cannot assume that they are uniformly distributed in the respective address segment. Specifically, the number of distinct edge label is limited by the number of unique ``verbs'' used in describing the stored information. To achieve a more uniform distribution, we are ``hashing'' the tuple by bit-interleaving the individual components and use that as the tuple's address.

\begin{gather*}
[[a_{127}, a_{126}, ..., a_{0}], \; [b_{127}, ..., b_{0}], \; [c_{127}, ... , c_{0}]] \\
\Downarrow \\
[a_{127}, b_{127}, c_{127}, \; a_{126}, b_{126}, c_{126}, ..., \; a_{0}, b_{0}, c_{0}]
\end{gather*}

Although we have not mathematically verified that this indeed leads to a more uniform distribution across the available address space, an analysis of our experiments do indeed show such characteristics.

The only other modification we have made  to support the storage of tuples was to increase the width of the address space from the traditional 160 to 512 bits and picked a ``non-value'' symbol as tuples with only labeled nodes do not have a value to store. While the increase in address space does increase the message size between the Kademlia peers, we developed some new data structures and lazy mechanisms to instantiate them to minimize its impact on the nodes memory consumption. In practice, the effective address width is $log_{2}(N) + k$ with $N$ being the number of peers and a very small $k$ accounting for the variations in the peer's address distribution.

Larger changes are necessary to support the {\em subscribe} operation. At this stage, we have implemented the standing query nature of subscribe as a periodic query on the Kademlia network and only support filter templates where each element either holds a specific label, or a wildcard which matches anything. We then translate the filter into a lookup vector which is reshuffled the same way as described above, however now every element can contain either a 0, 1, or $*$ (``don't care'').  The modified lookup algorithm now needs to branch whenever it encounters a $*$, with one branch substituting a 0 and the other a 1. Similar to the store operation, the number of branching points is related to the number of wildcards in the first $log_{2}(N) + k$ bits with the remaining part taken care by wildcard enabled lookup on a node's internal store, which not surprisingly, we also implemented using a tree structure.

\label{prototype}
\section{Conclusion and Future Work}\label{conclusion}

We have attempted to make a case for global information network which we believe is well suited for many of the current and emerging services as Metcalfe's law is even more relevant in the information realm. 

We discussed a set of design goals, described an architecture realization and provided an overview of our prototyping efforts to validate various systems aspects. This includes applications, higher-level subscription languages, and an implementations of the network service.

Future work will concentrate on developing an appropriate evaluation regime for this type of network, as well as a better understanding of the trust and security aspects of a global information network.

%



\label{conclusion}

\newpage
\bibliographystyle{IEEEtran}
\bibliography{references}

\end{document}